\documentclass[aps,twocolumn,showpacs,groupedaddress]{revtex4}
\usepackage{amssymb}
\usepackage{graphicx}
\usepackage{amsmath}
\newcommand{\Tr}{\text{Tr}}
\newcommand{\ket}[1]{|#1\rangle}
\newcommand{\bra}[1]{\langle#1|}
\begin{document}

\title{Singularity of dynamical maps}
\author{S. C. Hou$^1$, X. X. Yi$^{1,2}$, S. X. Yu$^{2,3}$ and C. H. Oh$^2$}
\affiliation{$^1$School of Physics and Optoelectronic Technology\\
Dalian University of Technology, Dalian 116024 China\\
$^2$Centre for Quantum Technologies and Department of Physics,
National University of Singapore, 117543, Singapore \\
$^3$Hefei National Laboratory for Physical Sciences at Microscale
and Department of Modern Physics, University of Science and
Technology of China, Hefei, 230026 Anhui, China}

\date{\today}

\begin{abstract}
For a dynamical map $\Lambda(t,0)$, which sends a state $\rho(0)$ of
quantum open system to a state $\rho(t)=\Lambda(t,0)\rho(0)$, the
decomposition law $\Lambda(t,0)=\Lambda(t,t_c)\Lambda(t_c,0)$ may
break down at a specific  time $t_c$. In this paper, we present a
method to find the singular points $t_c$ and propose a measure for
the singularity of the dynamical map. Two examples are portrayed  to
illustrate the method, the measure of singularity for these singular
points is calculated and discussed. An extension to high-dimensional
system is presented.
\end{abstract}

\maketitle

\section{introduction}
The actual dynamics of any real open quantum system is expected to
deviate to some extent from the Markovian evolution. This deviation
can be measured by non-Markovianity and it has attracted much
attention in recent years, leading to a deeper understanding of
quite a few issues in the theory of open quantum system
\cite{stockburger02,daffer04,piilo08,
breuer08,shabani09,chruscinski10,barchielli10,laine10}.

Non-Markovian systems can be found  in many branches of physics,
including  quantum optics \cite{breuer07,gardiner99}, solid state
physics \cite{lai06}, quantum chemistry\cite{plenio98}, and quantum
information processing \cite{aharnov06}. Since non-Markovian
dynamics modifies monotonic decay of quantum coherence, it may
protect quantum entanglement  in composite systems for longer time
than standard Markovian evolution \cite{nski10}. In particular it
may protect the system against the sudden death of entanglement
\cite{yu07}. Therefore, it is  interesting to quantify the
non-Markovianity within the description of quantum open system.

There are two approaches to quantify  the measure of the degree of
non-Markovianity. One approach is based on the idea of the
composition law which is essentially equivalent to the idea of
divisibility \cite{wolf08}. This approach was used recently in
Ref.\cite{rivas10,yi11} to construct the  measure of
non-Markovianity, quantifying  actually the deviation of the
dynamical map from divisibility. Another  approach is as  in
Ref.\cite{breuer09}, where the authors define non-Markovianity
dynamics as the information flow from the environment back into the
system, the measure manifests itself as an increase in the
distinguishability of pairs of evolving quantum states, and the
information is identified to be the Fisher information \cite{lu10}.

The measure for non-Markovianity proposed in Ref.\cite{rivas10} is
based on the completely positive divisibility of a dynamical map: a
trace preserving completely positive map $\Lambda(t,0)$ is
completely positive divisible (CP-divisibility) if it can be written
as,
\begin{eqnarray}
\Lambda(t_2,0)=\Lambda(t_2,t_1)\Lambda(t_1,0),
\end{eqnarray}
and $\Lambda(t_2,t_1)$ is  completely positive for any $t_2$ and
$t_1$ ($t_2>t_1>0$). By contrast, we say that the map
$\Lambda(t_2,0)$ is positively divisible (P-divisibility) if
$\Lambda(t_2,t_1)$ sends states into states but it is only positive,
and that $\Lambda(t_2,0)$ is indivisible if neither P-divisibility
nor CP-divisibility holds.

In this paper, we shall consider the other situation where,
\begin{eqnarray}
\Lambda(t_2,0)\neq \Lambda(t_2,t_c)\Lambda(t_c,0),
\end{eqnarray}
at a special time $t_c$, $t_2<t_c<0$. We will refer to this instance
of time $t_c$ as the singular point of the dynamical map
$\Lambda(t_2,0), t_2\in (0,\infty).$ Taking a qubit (two-level
system) as an example, a method to find the singular point is
presented, a measure to quantify this singularity is proposed and
discussed.

This paper is organized as follows. In Sec.{\rm II}, we present a
general formalism for a qubit dynamics, exhibiting  the method to
find the singular point $t_c$. A measure to quantify the singularity
is constructed.  Two examples, one describes a qubit coupled to a
harmonic oscillator  bath and the other  includes a qubit coupled to
a finite spin bath, are given to illustrate the critical point in
Sec.{\rm III},   the measure of singularity is also calculated and
discussed in this section. A generalization of the representation to
$d$-dimensional open systems is presented in Sec. {\rm IV}. Finally,
we conclude our results in Sec. {\rm V}.

\section{general formalism for a qubit dynamical map}
Consider a dynamical map $\Lambda(t,0)$  for a qubit (or two-level
system), which sends an arbitrary initial state
$\rho(0)=(1+\vec{n}(0)\cdot \vec{\sigma})/2$ with Bloch vector
$\vec{n}(0)=(n_x(0),n_y(0),n_z(0))$ into a state $\rho(t)$,
\begin{eqnarray}
\rho(t)=\Lambda(t,0)\rho(0)=\frac{1}{2}(1+\vec{n}(t)\cdot\vec\sigma).
\end{eqnarray}
Without loss of generality, the Bloch vector
$\vec{n}(t)=(n_x(t),n_y(t),n_z(t))$ can be written as,
\begin{equation}
\vec{n}(t)=\vec{n}(0)\cdot D(t)+\vec{f}(t), \label{bloch}
\end{equation}
where $D(t)$ is a $3\times 3$ matrix and $\vec{f}(t)$ is a
time-dependent vector.

Now we elicit   the condition for
$\Lambda(t_2,0)\neq\Lambda(t_2,t_c)\Lambda(t_c,0)$. To this aim we
introduce an ancilla $A$ and define,
\begin{eqnarray}
M_{SA}\equiv\Lambda(t_2,t_c)\otimes
\textrm{I}_A(\ket{\Phi_{SA}}\bra{\Phi_{SA}}),
\end{eqnarray}
where
$\ket{\Phi_{SA}}=\ket{0}_S\otimes\ket{0}_A+\ket{1}_S\otimes\ket{1}_A$
is an unnormalized  maximally entangled state of the qubit and
ancilla, and $\textrm{I}_A$ denotes the identity operator  of the
ancilla. Since the ancilla is also a qubit, $M_{SA}$ can be written
as,
\begin{eqnarray}
M_{SA}&=&\frac{1}{2}(x+\vec{r}\cdot\vec{\sigma}_A^T+ \vec{s}\cdot
\vec{\sigma}_S+\vec{\sigma}_S\cdot V\cdot
\vec{\sigma}_A^T)\nonumber\\
&=&\frac{1}{2}(\textrm{I},\vec{\sigma}_S)F \left(
\begin{array}{c}
               1 \\
\vec{\sigma}_A^T \\
\end{array}
\right).
\end{eqnarray}
Here $x$ is a constant, $\vec{r}$ and $\vec{s}$ are vectors,
$\textrm{I}$ is an identity matrix,  $F=\left(
    \begin{array}{cc}
      x & \vec{r} \\
      \vec{s} & V \\
    \end{array}
  \right)$ and $V$ is  a $3 \times 3$ matrix, which is
determined by the map $\Lambda(t_2,t_c)$ and will be derived  in the
following. If $\Lambda(t_2,0)=\Lambda(t_2,t_c)\Lambda(t_c,0)$ holds,
the map $\Lambda(t_2,t_c)$ would send the state
$\rho(t_c)=\Lambda(t_c,0)\rho(0)$ to state $\rho(t_2)$. In terms of
$M_{SA}$, this can be expressed as,
\begin{eqnarray}
\rho(t_2)&=&\Lambda(t_2,t_c)\rho(t_c)=\Tr_A[M_{SA}\
I_S\otimes\rho_A^T(t_c)]\nonumber\\
&=&\frac{1}{2}(\textrm{I},\vec{\sigma}_S)F \left(
\begin{array}{c}
               1 \\
\vec{n}(t_c) \\
\end{array}
\right). \label{rhot2}
\end{eqnarray}
Writing $\rho(t_2)=\frac 1 2
(\textrm{I},\vec{\sigma}_S)\left(\begin{array}{c}
               1 \\
\vec{n}(t_2) \\
\end{array}
\right),$ we obtain from Eq.(\ref{rhot2}),
\begin{eqnarray}
 \left(
\begin{array}{c}
               1 \\
\vec{n}(t_2) \\
\end{array}
\right)=F
\left(
\begin{array}{c}
               1 \\
\vec{n}(t_c) \\
\end{array}
\right)=
\left(
\begin{array}{c}
x+\vec{r}\cdot\vec{n}(t_c) \\
\vec{s}+ \vec{n}(t_c)\cdot V \\
\end{array}
\right).
\end{eqnarray}
It is easy to find that,
\begin{eqnarray}
x=1,\ \ \ \vec{r}=0,\nonumber\\
\vec{n}(t_2)=\vec{n}(t_c)\cdot V+\vec{s}.\label{bloch1}
\end{eqnarray}
Considering that $\rho(0)$ is an arbitrary initial state, namely
$\vec{n}(0)$ is arbitrary,  Eqs.(\ref{bloch},\ref{bloch1}) together
yield,
\begin{eqnarray}
D(t_2)=D(t_c)\cdot V, \nonumber\\
\vec{f}(t_2)=\vec{f}(t_c)\cdot V+\vec{s}. \label{Newc}
\end{eqnarray}
The condition for $\Lambda(t_2,0)\neq\Lambda(t_2,t_c)\Lambda(t_c,0)$
now is equivalent to that there does not exist a matrix $V$ to
satisfy Eq. (\ref{Newc}). If the determinant of  $D(t_c)$ is
non-zero,  we have
\begin{equation}
V=D^{-1}(t_c)\cdot D(t_2),
\end{equation}
and
\begin{eqnarray}
\vec{s}=\vec{s}(t_2,t_c)=\vec{f}(t_2)-\vec{f}(t_c)\cdot V.
\end{eqnarray}
From the above derivations, we find that once $x$, $\vec{r}$,
$\vec{s}$, $V$ are (uniquely or non-uniquely)  established for any
$t_c$ ($t_2>t_c>0$), the decomposition
$\Lambda(t_2,0)=\Lambda(t_2,t_c) \Lambda(t_c,0)$ holds true, namely
there are no singular points in the time interval $[0,t_2].$ We
notice that the null-determinant of $D(t_c)$ plays a key role in
finding $V$, it can  thus be taken as a condition to find the
singular point $t_c$ if the matrix $D(t_2)$ is of full rank.
Mathematically,  the necessary and sufficient condition for Eq.
(\ref{Newc}) to have no solution is that the rank of $D(t_c)$ must
be smaller than the rank of $[D(t_c)|D(t_2)]$, where
$[D(t_c)|D(t_2)]$ is an augmented matrix obtained by attaching the
columns of $D(t_2)$ to the columns of $D(t_c)$.

To quantify the singularity of the singular point, we introduce the
trace distance, $D(\rho_1,\rho_2)=\frac 1 2
\mbox{Tr}|\rho_1-\rho_2|,$ which is an appropriate measure for the
distinguishability between two quantum states $\rho_1$ and $\rho_2$.
Here $|A|=\sqrt{A^{\dagger}A}.$ We define the singularity measure of
a dynamical map $\Lambda(t_2,0)$ at time $t_c$ by
\begin{equation}
S_{\Lambda}(t_c)=\max_{\rho(0),T} D(\rho(T),\rho_{t_c}(T)),
\end{equation}
where $\rho(T)\equiv \Lambda(T,0)\rho(0)$ and $\rho_{t_c}(T)\equiv
\Lambda(T,t_c)\cdot \Lambda(t_c,0)\rho(0)$  is the solution of the
dynamical process taking $\rho(t_c)=\Lambda(t_c,0)\rho(0)$ as
initial state. The maximum is taken over all initial states and the
final time $T$.

\section{examples}
In this section, we will present two examples to illustrate the
singular point and the measure of singularity. The first example is
a dephasing model that consists of a spin-$\frac 1 2$ particle
coupling to a spin-bath. The coupling Hamiltonian commutes with the
free Hamiltonian of the central spin, thus the central spin
conserves its energy. In the second example, we consider a
dissipative system, the energy of the system is no longer conserved.
\subsection{A two-level system coupling to  a  finite spin bath}
Consider a central spin-$\frac 12$ coupling to a bath of $N$
spin-$\frac 1 2$ particles. The interaction Hamiltonian is,
\begin{equation}
H=\sum_{k=1}^NA_k\sigma_z\sigma_z^{k},
\end{equation}
where $A_k=A/\sqrt{N}$ represents the coupling constants.  Assume
the initial state of the whole  system is $\rho_s(0)\otimes
(\frac{1}{2^N}I)$, i.e.,  all  spins in the reservoir  are in  a
maximal mixed state. The density matrix of the central spin at time
$t$ takes,
\begin{equation}
\rho(t)= \left(
\begin{array}{cc}
\rho_{11} & \rho_{12}\cos^N(\frac{2At}{\sqrt{N}})\\
\rho_{21}\cos^N(\frac{2At}{\sqrt{N}}) & \rho_{22}\\
\end{array}
\right).\label{denc}
\end{equation}
In terms of dynamical map, the dynamics can be represented as,
$\Lambda(t,0)\rho=\frac 1 2 (1-\cos^N(\frac{2At}{\sqrt{N}}))
\sigma_z\rho\sigma_z+\frac{1}{2}(1+\cos^N(\frac{2At}{\sqrt{N}}))\rho.$
This is equivalent to  the following master equation,
\begin{equation}
\dot{\rho}=\gamma(t)\mathcal{L}(\rho), \label{eqn:tan}
\end{equation}
where $\mathcal{L}(\rho)=\sigma_z\rho\sigma_z-\rho$,   and the
time-dependent decay rate is
$\gamma(t)=A\sqrt{N}\tan(\frac{2At}{\sqrt{N}})$. This model is
discussed in several papers as a typical  example to quantify
non-Markovianity.

Writing $\rho(t)$ in Eq. (\ref{denc}) in the form of Eq.
(\ref{bloch}), we find
\begin{eqnarray}
D(t)&=&\left(
    \begin{array}{ccc}
      C(t) & 0 & 0 \\
      0 & C(t) & 0 \\
      0 & 0 & 1\\
    \end{array}
  \right),\nonumber\\
\vec{f}(t)&=&0,
\end{eqnarray}
where $C(t)=\cos^N(\frac{2At}{\sqrt{N}}).$ The singular point $t_c$
can be found by solving $C(t_c)=0,$ it yields,
\begin{equation}
t_c=\frac{\sqrt{N}}{4A}(2n+1)\pi, \ \ n=0,1,2,.... \label{tc1}
\end{equation}
By the definition of the measure of singularity, we obtain,
\begin{equation}
S_{\Lambda}(t_c)=\max_{|\rho_{12}|,T}|C(T)||\rho_{12}(0)|,
\end{equation}
where $T$ is a time, $T>t_c$, and $\rho_{12}(0)$ denotes the element
of the initial density matrix $\rho(0)=\left(
    \begin{array}{cc}
      \rho_{11}(0) & \rho_{12}(0)\\
      \rho_{21}(0) & \rho_{22}(0)\\
    \end{array}
  \right).$
After a simple algebra, we have $S_{\Lambda}(t_c)=\frac{1}{2}$ for
any singular point given in Eq.(\ref{tc1}). It is interesting  that
the singularity measure of these singular points are equal. Indeed,
examining the dynamical map $\Lambda(t,0)$, we find that the
features of $\Lambda(t,0)$ around any $t_c$ are the same.

\subsection{The damping J-C model}
This example consists of   a two-level system coupling to  a
reservoir at zero temperature. The reservoir consists of infinite
number of harmonic oscillators that  is also referred in the
literature as the spin-boson model.  The Hamiltonian for such a
system reads,
\begin{eqnarray}
H=H_0+H_I,
\end{eqnarray}
where $H_0=\hbar\omega_0\sigma_+\sigma_-
 +\sum_k\hbar\omega_k b_k^{\dag}b_k,$
$H_I=\sigma_+B+\sigma_-B^\dag,$ and  $B=\sum_k g_k b_k$. The Rabi
frequency  of the two-level system and the frequency for the $k-th$
harmonic oscillator  are denoted by $\omega_0$ and $\omega_k,$
respectively. $b^\dag_k$ and $b_k$ are the creation and annihilation
operators of  $k-th$ oscillator, which  couples to the system with
coupling constant $g_k$.

This model  is exactly solvable \cite{breuer07}. Assuming  the
system and the reservoir initially uncorrelated, we can obtain a
time-dependent master equation in the interaction picture,
\begin{eqnarray}
\dot{\rho}&=&-i\frac{e(t)}{2}[\sigma_+\sigma_-,\rho]\nonumber\\
&+&\gamma(t)(\sigma^-\rho\sigma^+-\frac{1}{2}\sigma^+
\sigma^-\rho-\frac{1}{2}\rho\sigma^+\sigma^-), \label{eqn:exactnm}
\end{eqnarray}
where $e(t)=-2\mathrm{Im}[\frac{\dot{c}(t)}{c(t)}]$  and
$\gamma(t)=-2\mathrm{Re}[\frac{\dot{c}(t)}{c(t)}]$. $e(t)$ plays the
role  of Lamb shift and $\gamma(t)$ is the decay rate. Both $e(t)$
and $\gamma(t)$ are  time-dependent. $c(t)$ is determined by
$\dot{c}(t)=-\int_0^t f(t-\tau)c(\tau)d(\tau)$, where
$f(t-\tau)=\int d\omega J(\omega) exp(i(\omega_0-\omega)(t-\tau))$
is the environmental correlation function. In the derivation of the
master equation, the reservoir is assumed in its vacuum at $t=0$.

Consider the following spectral density, $
J(\omega)=\frac{1}{\pi}\frac{\gamma_0\lambda^2}
{(\omega_0-\omega)^2+\lambda^2},$  where $\gamma_0$ represents the
coupling constant between the system and reservoir, $\lambda$
defines the spectral width of  the coupling at the resonance point
$\omega_0$. For the spectral density $J(\omega)$, we have  $e(t)=0$,
$c(t)=c_0 e^{-\lambda
t/2}[\cosh(\frac{dt}{2})+\frac{\lambda}{d}\sinh(\frac{dt}{2})]$, and
\begin{equation}
\gamma(t)=\frac{2\gamma_0\lambda
\sinh(dt/2)}{d\cosh(dt/2)+\lambda\sinh(dt/2)}
\end{equation} with
$d=\sqrt{\lambda^2-2\gamma_0\lambda}$ in Eq.(\ref{eqn:exactnm}).
Assume the system initially in  $\rho(0)=\left(
    \begin{array}{cc}
      \rho_{ee}(0) & \rho_{eg}(0)\\
      \rho_{ge}(0) & \rho_{gg}(0)\\
    \end{array}
  \right),$
by the effective Hamiltonian approach\cite{yi01}, we have the
density matrix at time $t$, $\rho(t)=\left(
    \begin{array}{cc}
      \rho_{ee}(t) & \rho_{eg}(t)\\
      \rho_{ge}(t) & \rho_{gg}(t)\\
    \end{array}
  \right),$
where
\begin{eqnarray}
\rho_{ee}(t)&=&\rho_{ee}(0)e^{-\int_0^t\gamma(t')dt'},\nonumber\\
\rho_{gg}(t)&=&1-\rho_{ee}(t),\nonumber\\
\rho_{eg}(t)&=&\rho_{ge}^*(t)=e^{-\frac 1
2\int_0^t\gamma(t')dt'}\rho_{eg}(0).
\end{eqnarray}
It is easy to show that the matrix $D(t)$ and $\vec{f}(t)$ in this
example are,
\begin{eqnarray}
 D(t)=\left(
    \begin{array}{ccc}
      D_{11}(t) & 0 & 0 \\
      0 & D_{22}(t) & 0 \\
      0 & 0 & D_{33}(t) \\
    \end{array}
  \right) \label{Deq}
\end{eqnarray}
and
\begin{eqnarray}
\vec{f}(t)=(f_x, f_y, f_z)=(0,0,(e^{-\int_0^t\gamma(t')dt'}-1)),
\end{eqnarray}
where
\begin{eqnarray}
D_{11}&=&D_{22}=e^{-\frac{1}{2}\int_0^t\gamma(t')dt'},\nonumber\\
D_{33}&=&D_{11}^2.
\end{eqnarray}
We find from Eq. (\ref{Deq}) that   $D_{jj}(t_c)=0,$ $j=1,2,$ or
$3,$ gives  the singular points.   $D_{jj}(t_c)=0$ can happen only
when $\gamma_0/\lambda>1/2.$ Noticing that $D_{11}(t)=e^{-\lambda
t/2}[\cos(\frac{d_0t}{2})+\frac{\lambda}{d_0}\sin(\frac{d_0t}{2})]$
for $\gamma_0/\lambda>1/2,$ where
$d_0=\sqrt{|\lambda^2-2\gamma_0\lambda|},$ we obtain the $n$th
singular point $t_c^{(n)}=$
$\frac{2}{d_0}(\cot^{-1}(-\frac{1}{\sqrt{|1-2\gamma_0/\lambda|}})+n\pi),n=0,1,2,....$
At these singular points, the singularity measure  can be given by
maximizing the distance $ D(\rho(T),\rho_{t_c}(T))
=\frac{1}{2}\sqrt{D_{33}(T)(n_1^2(0)+n_2^2(0))+D_{33}^2(T)(1+n_3(0))^2}$
over $T$ and the Bloch vector $\vec{n}(0)=(n_1(0),n_2(0),n_3(0))$
with constraint  $0\leq n_1^2(0)+n_2^2(0)+n_3^2(0)\leq 1.$ Simple
algebra shows that the  maximum of the $n$th singular point arrives
at $T=T^{(n)}=\frac{2(n+1)\pi}{d_0},\, n=0,1,2...,$ and
$n_3(0)=\frac{D_{33}(T^{(n)})}{1-D_{33}(T^{(n)})},
n_1^2(0)+n_2^2(0)=1-n_3^2(0).$ The measure of singularity for the
$n$th singular point $t_c^{(n)}$ is then
\begin{eqnarray}
S_{\Lambda}(t_c^{(n)})&=&\frac{1}{2}\sqrt{\frac{
e^{-\frac{2(n+1)\pi\lambda}
{d_0}}}{{1-e^{-\frac{2(n+1)\pi\lambda}{d_0}}}}},\,\, \mbox{for} \,
\,
0\leq D_{33}(T^{(n)})< 0.5. \nonumber\\
S_{\Lambda}(t_c^{(n)})&=&e^{-\frac{2(n+1)\pi\lambda}{d_0}},\,\,
\mbox{for}\,\,  0.5\leq D_{33}(T^{(n)})\leq 1.
\end{eqnarray}

We find that the values of singularity measures  are different in
the two examples. In the first example, the singularity for all
singular points are the same,  $S_{\Lambda}(t_c)=\frac 1 2 $, while
in the second one, the singularity depends on the singular points.
This results from the difference in the states at the singular point
$t_c$ in the two examples. Especially, as $n$ increases, the
singularity decreases and finally tends to zero as $n\rightarrow
\infty.$ In other words, the singularity of larger $t_c$ is smaller
than that for a smaller $t_c.$ This can be understood as that at
large $t_c,$ the state of the open system is more close to the
steady state, leading to a small difference in the states.
Furthermore, the difference in singularity is a reflection of the
system-environment coupling. The first example is a dephasing model,
it conserves the system energy but spoils the off-diagonal elements
of the density matrix. By contrast, the system would decay to its
ground state at the singular points in the second example.

Since the measure proposed here quantifies the non-divisibility of
the dynamical map, $\Lambda(t,0)\neq \Lambda(t,t_c)\Lambda(t_c,0),$
hence it can measure the non-Markovianity of the map. The
non-Markovianity in this situation  depends both on the measure of
the singularity and the number of singular points. Therefore, we
propose,
\begin{equation}
N_M=\sum_j S_{\Lambda}(t_{c}^{(j)}),
\end{equation}
to quantify the non-Markovianity caused by the singular points
$t_{c}^{(j)}, \, (j=1,2,3,...).$ Physically, once a dynamical map
has a singular point $t_c$, the state at time $t>t_c$ would depend
on the state at an earlier time $t^{\prime} <t_c$, although the
state at $t_c$ is the same. This feature can be found by examining
Eq.(\ref{denc}), which is a reminiscence of the classical
non-Markovian process.

The present prediction can be observed in the experimental setup in
\cite{liu11}, where the polarization degree of freedom of photons
plays the role of open system, the environment was simulated by the
frequency degree of freedom with two central frequencies at
$\omega_1$ and $\omega_2$. The evolution of the off-diagonal
elements of the photon density matrix takes, $|H\rangle\langle
V|\rightarrow \kappa^*(t) |H\rangle\langle V|, \,\,$$
|V\rangle\langle H|\rightarrow \kappa(t) |V\rangle\langle H|.$ Here
$\kappa(t)$ is adjustable and can be manipulated to  zero at times
$t_c=-\frac{(2n+1)\pi}{\Delta \omega\cdot \Delta n}, n=0,1,2,...,$
where $\Delta\omega=\omega_2-\omega_1,$ is the difference in the
central frequencies of the environment. $\Delta n$ denotes the
difference in the refraction indices of horizontally and vertically
polarized photons. The observed final states are different that
depend on whether an observation is made at the singular points
$t_c$. By measuring the difference in the final states, the
singularity can be quantified in the experiment.

\section{extension to $d$-dimensional systems}
We consider now an arbitrary dynamical  map $\Lambda(t,0)$ with
$t>0$ for a quantum $d$-dimensional system (qudit). Let
$\{\lambda_\mu\}_{\mu=1}^{n}$ with $n=d^2-1$ be a set of traceless
qudit observable  satisfying
$\Tr(\lambda_\mu\lambda_\nu)=d\delta_{\mu\nu}$. Together with the
identity operator they form an orthonormal basis for all the qudit
operators. Thus we have expansions $\varrho=(I+\vec n\cdot\vec
\lambda)/d$ for the initial state and $\Lambda(t,0)\varrho=(I+\vec
n(t)\cdot\vec\lambda)/d$ with $\vec n(t)= D(t)\cdot\vec n+\vec e(t)$
for the final state $\varrho(t)=\Lambda(t,0) \varrho$. Here $\vec
n=\Tr(\varrho \vec\lambda)$, $\vec n(t)=\Tr(\varrho(t)
\vec\lambda)$, and $\vec e(t)=\Tr[(\Lambda(t,0) I) \vec\lambda]$ are
$n$-dimensional real vectors and $D(t)$ is an $n\times n$ real
matrix with matrix elements given by
$[[D(t)]]_{\mu\nu}=\Tr((\Lambda(t,0) \lambda_\nu)\lambda_\mu).$ The
linear trace-preserving map $\Lambda(t,0)$ is determined uniquely by
$D(t)$ and $\vec e(t)$ and vice versa. For later use we denote by
$V(t)=\{\vec a\in R^n|D(t)\cdot \vec a=0\}$ the null space of
$D(t)$.

Let $t>t_c>0$ and consider the possible decomposition of a dynamical
map $\Lambda(t,0)=\Lambda(t,t_c)\Lambda(t_c,0)$  for some linear
trace-preserving map $\Lambda(t,t_c)$.  Any linear qudit map, e.g.,
$\Lambda(t,t_c)$, is in a one-to-one correspondence with a 2-qudit
operator, e.g., $R=\Lambda(t,t_c)\otimes \mathcal I(\Phi)$ with
$\Phi$ being the projector of the subnormalized 2-qudit state
$|\Phi\rangle=\sum_{n}|n,n\rangle$. We denote by $S$ the $n\times n$
matrix with elements $[[S]]_{\mu\nu}=\Tr
(R\lambda_\mu\otimes\lambda_\nu^T)/d$ for
$\mu,\nu=1,2,\ldots,n=d^2-1$ and $\vec r=\Tr (R\vec \lambda\otimes
I)/d$. For a trace-preserving map it holds $\Tr (RI\otimes \vec
\lambda)=0$ and therefore $S$ and $\vec r$ determine uniquely $R$
and consequently the linear trace-preserving map $\Lambda(t,t_c)$.
By definition the linear map $\Lambda(t,t_c)$ is a possible
decomposition if and only if for an arbitrary initial state
$\varrho$ it holds
\begin{equation}\label{cond}
\varrho(t)=\Lambda(t,0)\varrho=\Lambda(t,t_c)(\Lambda(t_c,0)
\varrho) =\Lambda(t,t_c)\varrho(t_c).
\end{equation}

{\it Lemma. } If a qudit operator $O$  satisfies $\langle \psi
|O|\psi\rangle=0$ for an arbitrary pure qudit state $|\psi\rangle$
then $O=0$.

{\it Proof. } Let $V_{12}=\sum_{ij}|i,j\rangle\langle j,i|$ be the
swapping operator of 2 qudits and $I_{12}$ be the identity operator.
From the following identity
\begin{equation}
W_{12}:=\int d\psi |\psi\rangle\langle\psi|\otimes
|\psi\rangle\langle\psi|=\frac{I_{12}+V_{12}}{d(d+1)}
\end{equation}
it follows that $O=d(d+1)\Tr_1 \big((O_1\otimes I_2) W_{12}\big)=0$.

{\it Theorem. } Given a qudit channel $\Lambda(t,0)$ with $t> t_c>0$
there exists a linear trace-preserving map $\Lambda(t,t_c)$ such
that $\Lambda(t,0)=\Lambda(t,t_c) \Lambda(t_c,0)$ if and only if
$V(t_c)\subseteq V(t)$. Moreover the decomposition $\Lambda(t,t_c)$
is unique if and only if $\det D(t_c)\not=0$.

{\it Proof.} Necessity (only if part), i.e.,
$\Lambda(t,0)=\Lambda(t,t_c) \Lambda(t_c,0)$ infers $V(t_c)\subseteq
V(t)$. From Eq.(\ref{cond}) it follows that for arbitrary $\varrho$
it holds
 $\Tr(\vec \lambda\varrho(t))= \Tr\big(\vec \lambda\Lambda(t,t_c)
 \varrho(t_c))\big)=\Tr (R\vec\lambda\otimes \varrho^T(t_c))=
 S\cdot\vec n(t_c)+\vec r$. Taking into account $\vec n(t)=
 \Tr(\vec \lambda\varrho(t))=D(t)\cdot \vec n+\vec e(t)$ for $t$ and $t_c$ we see that
$(D(t)-S\cdot D(t_c))\cdot \vec n=S\cdot \vec e(t_c)+\vec r-\vec
e(t)$ must hold for arbitrary $\vec n=\Tr(\varrho\vec \lambda)$ with
$\varrho$ being a density matrix. If we let $\vec n=0$ with
corresponding state being $\varrho=I/d$ then we obtain $\vec r=\vec
e(t)-S\cdot \vec e(t_c)$. As a result $\Delta\cdot \vec n=0$, where
$\Delta:=D(t)-S\cdot D(t_c)$, for arbitrary $\vec n=\Tr(\varrho\vec
\lambda)$ with $\varrho$ being a density matrix.
 From $\Delta\cdot\Tr (\varrho \vec \lambda)=0$ it follows that
 $\Tr(\varrho \Delta_\mu)=0$ for arbitrary $\mu$ and arbitrary
 qudit state $\varrho$ where $\Delta_\mu=\sum_\nu \Delta_{\mu\nu}\lambda_\nu$
 is traceless, i.e.,  $\Tr\Delta_\mu=0$.
 As a result of lemma, $\Delta_\mu=0$ for all $\mu$, i.e., $D(t)=S\cdot D(t_c)$.
 Thus we have $V(t_c)\subseteq V(t)$ since $D(t_c)\cdot\vec n=0$
 infers $D(t)\cdot\vec n=S\cdot D(t_c)\cdot\vec n=0$.

Sufficiency (if part), i.e.,  $V(t_c)\subseteq V(t)$ infers
$\Lambda(t,0)=\Lambda(t,t_c) \Lambda(t_c,0)$. Let $\{\vec
e_i\}_{i=1}^{K}$ span $V(t_c)$ where $K=\dim V(t_c)$ and $\{\vec
e_{i}\}_{i=K+1}^n$ span the orthogonal complement of $V(t_c)$. As a
result $D(t_c)\cdot \vec e_i=0$ and thus $D(t)\cdot \vec e_i=0$ for
$i=1,2,\ldots, K$ since we have assumed $V(t_c)\subseteq V(t)$.
 The equation $D(t)=S\cdot D(t_c)$ is equivalent to
 $D(t)\cdot\vec e_i=S\cdot D(t_c)\cdot\vec e_i$ for $i=1,\ldots n$.
 Since $D(t_c)\cdot \vec e_i=D(t)\cdot \vec e_i=0$ for $i=1,\ldots, K$
 the equation becomes $D^\prime(t)=S\cdot D^\prime(t_c)$ with
 $D^\prime(t_c)=[D(t_c)\vec e_{K+1},D(t_c)\vec e_{K+2},\ldots D(t_c)\vec e_{n}]$
 and $D^\prime(t)=[D(t)\vec e_{K+1},D(t)\vec e_{K+2},\ldots,D(t)\vec e_{n}]$
 being of dimension $n\times (n-K)$.
Since the rank of $D(t_c)$ is $n-K$ there are exactly $n-K$ (among
$n$) linearly independent row vectors of $D^\prime(t_c)$. Therefore
it is always possible to expand each row vector of $D^\prime(t)$, an
$(n-K)$-dimensional vector,  by  those $n$ row vectors of
$D^\prime(t_c)$, i.e., for any given $i$ there exist real numbers
$S_{ik}$ such that
\begin{eqnarray}
([D^\prime(t)]_{i1},[D^\prime(t)]_{i2},\ldots,[D^\prime(t)]_{i,n-K})=\hskip
2cm\cr \sum_{k=1}^nS_{ik}([D^\prime(t_c)]_{k1},[D^\prime(t_c)]_{k2},
\ldots,[D^\prime(t_c)]_{k,n-K}).
\end{eqnarray}
 The $n\times n$ matrix $S$ formed by the coefficients
 $S_{ik}$ with $i,k=1,2,\ldots,n$ satisfies $D(t)=S\cdot D(t_c)$
 and together with $\vec r=\vec e(t)-S\cdot \vec e(t_c)$ determines
 $R$ and thus $\Lambda(t,t_c)$ such that
 $\Lambda(t,0)=\Lambda(t,t_c)\Lambda(t_c,0)$.
 Moreover if and only if $K=0$, i.e., $V(t_c)$ is
empty,
 i.e, $\det D(t_c)\not=0$, $S$ as well as $\vec r=\vec e(t)-S\cdot \vec e(t_c)$
is unique and therefore $\Lambda(t,t_c)$ is unique.

We note that the condition $V(t_c)\subseteq V(t)$ is equivalent to
$\mbox{Rank}[D(t_c)|D(t)]= \mbox{Rank}(D(t_c))$, where
$[D(t_c)|D(t)]$ is an augmented matrix. The decomposition
$\Lambda(t,0)=\Lambda(t,t_c)\Lambda(t_c,0)$ does not exist if and
only if $\mbox{Rank}[D(t_c)|D(t)]> \mbox{Rank}(D(t_c))$.

\section{conclusion}
 In summary, we have explored the singular point $t_c$ where the
dynamical map $\Lambda(t,0)\neq\Lambda(t,t_c)\Lambda(t_c,0),$ i.e.,
the dynamical map is indivisible at the instance of time $t_c$. We
quantify the singularity  of the singular point $t_c$ and present
examples to show the singularity. Until now these points were not
aware in the divisibility-based measure of non-Markovianity, hence
it would contribute to the understanding of quantum non-Markovian
process.

Special thanks to Prof J.I. Cirac for his comments on the
divisibility-based measure of non-Markovianity, which  motivate the
present study. This work is supported by the NSF of China under
Grants No 61078011, No 10935010, and No 11175032 as well as the
National Research Foundation and Ministry of Education, Singapore,
under academic research grant No. WBS: R-710-000-008-271.

\end{document}